\documentclass{svmult}

\usepackage{makeidx}     % allows index generation
\usepackage{graphicx}    % standard LaTeX graphics tool
\usepackage{multicol}    % used for the two-column index
\makeindex             % used for the subject index
\begin{document}

\title*{Statistical properties of absolute log-returns and a stochastic model of stock markets with heterogeneous agents}
\titlerunning{Absolute log-returns and a stochastic model of stock markets} 
\author{Taisei Kaizoji}
\authorrunning{T. Kaizoji}
\institute{1. Division of Social Sciences, International Christian University 
3-10-2 Osawa, Mitaka, Tokyo 181-8585, Japan. \\ 
2. Econophysics Laboratory, 5-9-7-B Higashi-cho, Koganei-shi, Tokyo 184-0011, Japan. \texttt{kaizoji@icu.ac.jp, http://subsite.icu.ac.jp/people/kaizoji/}}
%
% Use the package "url.sty" to avoid
% problems with special characters
% used in your e-mail or web address
%
\maketitle

\begin{abstract} 
This paper is intended as an investigation of the statistical properties of {\it absolute log-returns}, defined as the absolute value of the logarithmic price change, for the Nikkei 225 index in the 28-year period from January 4, 1975 to December 30, 2002. We divided the time series of the Nikkei 225 index into two periods, an inflationary period and a deflationary period. We have previously [18] found that the distribution of absolute log-returns can be approximated by the power-law distribution in the inflationary period, while the distribution of absolute log-returns is well described by the exponential distribution in the deflationary period.\par
To further explore these empirical findings, we have introduced a model of stock markets which was proposed in [19,20]. In this model, the stock market is composed of two groups of traders: {\it the fundamentalists}, who believe that the asset price will return to the fundamental price, and {\it the interacting traders}, who can be noise traders. We show through numerical simulation of the model that when the number of interacting traders is greater than the number of fundamentalists, the power-law distribution of absolute log-returns is generated by the interacting traders' herd behavior, and, inversely, when the number of fundamentalists is greater than the number of interacting traders, the exponential distribution of absolute log-returns is generated. \par
\end{abstract}

\section{Introduction}
Over the past few decades a considerable number of studies have been made of distributions of financial data. In the early studies, studies of daily fluctuations of commodity prices led Mandelbrot \cite{1} and Fama \cite{2} to propose that the associated distribution function is a power-law distribution with an exponent close to $ 1.7$. More recently, some researchers have quantitatively reinvestigated the tick-by-tick data set of stock prices, which includes approximately $40$ million records, and found that the probability distribution of log-returns and volatility satisfies a power law with an exponent close to $3$ [3-6]. Theorists also use various versions of multiplicative random processes in the theoretical modeling of returns and volatility. An important goal of these market models is to explain the main stylized facts such as power laws observed in financial markets [7-17]. \par
In this present study, we first examined the statistical properties of absolute log-returns, defined as the absolute value of the logarithmic price change, for the Nikkei 225 index. We divided the time series of the index into two periods, an inflationary period and a deflationary period. Using this division, we found that the distribution of the absolute log-returns can be approximated by the power-law distribution in the inflationary period, while the distribution of absolute log-returns is well-described by the exponential distribution in the deflationary period\footnote{We have previously reported these empirical findings [18] and analyzed the results from a slightly different perspective.}. \par
We then performed numerical simulations of the model of stock markets proposed in our previous works [19,20] in order to provide an explanation of the empirical findings. In our model, the stock market is composed of two typical groups of traders: the fundamentalists who believe that the asset price will return to the fundamental price, and the interacting traders who tend to be influenced by the investment attitudes of other traders. We show that our model can reproduce the above-mentioned statistical properties of absolute log-returns. In particular, we show the following: 
\begin{itemize}
\item When (i) the number of interacting traders is greater than the number of fundamentalists, and (ii) the interaction among interacting traders is strong, a power-law distribution of absolute log-returns is generated. 
\item Inversely, when (i) the number of fundamentalists is greater than the number of interacting traders, and (ii) the interaction among interacting traders is weak, the exponential distribution of absolute log-returns is generated. 
\end{itemize}
Thus the numerical simulation helps account for the empirical findings. \par
The rest of the paper is organized as follows: the next section reviews briefly the empirical findings. Section 3 describes the stochastic model, section 4 provides the results of the numerical simulation, and section 5 gives concluding remarks. 

\section{The Empirical Results [18]}
\noindent
In this section we discuss the statistical properties of the absolute log-returns for the Nikkei 225 index in the 28-year period from January 4, 1975 to December 30, 2002. Figure 1(a) shows the time series of the Nikkei 225 index throughout the entire period. The Nikkei 225 index reached a high of almost 40,000 yen on the last trading day of the decade of the 1980s. However, the index had declined to 14,309 yen, a drop of approximately 63 percent by mid-August 1992. It follows from the figure that we can divide the time series of the Nikkei 225 index into two periods, an inflationary period from January 4, 1975 to December 29, 1989, and a deflationary period from January 4, 1990 to December 29, 2002. We define the return $ R_j $ as the change in the logarithm of the index $ R_j = ln S_j - ln S_{j-1} $, where $ S_j $ denotes the index at data $ j $. The term volatility represents a generic measure of the magnitude of market fluctuations. Thus, many different quantitative definitions of volatility are used in the literature. In this study, we define the volatility as the absolute value of the return, $ V_j = |R_j| $. Figure 1(b) shows the time series of the absolute log-return $ V_j $ of the Nikkei 225 index. \par
Figure 2(a) and 2(b) show the semi-log plots of the complementary cumulative distribution function of the absolute log-returns for the Nikkei 225 index in the inflationary and deflationary periods, respectively. 
We found that the complementary cumulative distribution function of the absolute log-returns for the deflationary period is accurately described by an exponential distribution $ P(V \geq x) \sim exp(-x/\beta) $ in the whole range of the absolute log-returns $ V_j $. We estimate the parameter $ \beta$ using the least-square method, and we estimate, 
$$ \beta = 0.68, \quad R^2 = 0.996, \quad \mbox{p-value}= (0.0). $$
In contrast, the other hand the complementary cumulative distribution function of the absolute log-returns for the inflationary period fits the exponential law for low values of $ x $, and it deviates from the exponential distribution as the absolute log-return increases. Figure 3 shows the log-log plot of the complementary cumulative distribution function of the absolute log-returns in the inflationary period. It is apparent that this plot tends to a linear function for high values of $ x $, indicating  that the complementary cumulative distribution function of the absolute log-returns is consistent with power law asymptotic behavior $ P(V \geq x) \sim x^{-\alpha} $. The value of $ \alpha $ estimated by the least squared method in $ [1.1 \leq x \leq 8] $, 
$$ \alpha = 3.38, \quad R^2 = 0.991, \quad \mbox{p-value} = (0.0). $$
In summary, we find that the distribution of absolute log-returns for stock price indices can be approximated by the power-law distribution in the inflationary period, while the distribution is accurately described by the exponential distribution in the deflationary period. \par
Why are empirically different laws observed?  To answer this question, we propose a stochastic model in the following section. 

\section{A Stochastic Model of Stock Markets [19,20]}
\noindent 
Let us consider a stock market where a stock is traded at price $ S_j $. Two groups of traders with different trading strategies, \textit{interacting traders} and, \textit{fundamentalists} participate in the trading. The number of fundamentalists $m$ and the number of interacting traders $n$ are assumed to be constant. The model is designed to describe the movements of the stock price over short periods, such as one day. In the following, a more precise account of the decision-making of each trader type is given. 

\subsection{Interacting traders}
\noindent 
During each time period, an interacting trader may choose to either buy or sell the index, and is assumed to trade a fixed amount of the index $b$ in a period of trading. Interacting traders are labeled by an integer $1 \leq i \leq n$. The investment attitude of the interacting trader $i$ is represented by the random variable $u_i$ and is defined as follows: If interacting trader $i$ is a buyer of the index during a given period, then $u_{i} = + 1$, otherwise he sells the index, and $u_{i} = - 1$. 
Now let us formulate the dynamics of the investment attitude of interacting traders. Consider that the investment attitude $u_{i}$ of interacting trader $i$ is updated with transition probabilities as a function of the average investment attitude over interacting traders $ X $ according to 
$ W_{\uparrow}(X) = 1/(1 + \exp(- 2 \phi X)) $ (transition from seller to buyer) and $ W_{\downarrow}(X) = 1/(1 + \exp(2 \phi X)) $ (transition from buyer to seller), where $ X = \sum^{n}_{i=1} u_i / n $ and $ \phi $ denotes the interaction strengths among traders. 
Using the above transition probabilities, the Ito stochastic differential equation for the average investment attitude $X$ is obtained as 
\begin{equation}
  d X(t) = K(X) dt + \sqrt{Q(X)} dW(t), 
\end{equation}
$$ K(X) = \tanh(\phi X) , \quad Q(X) = \frac{2}{n}(1 - \tanh(\phi X))  $$
where $ W(t) $ is the Wiener process. To perform the numerical simulation, we consider a discretized version of the stochastic differential equation, 
\begin{equation}
  X_{j+1} = X_j + K(X_j) \Delta t_j + \sqrt{Q(X_j)} \Delta W_j.
\end{equation}
Here, $ X_j = X(t_j) $, $ \Delta t_j = t_{j+1} - t_j $ and $ \Delta W_j = W(t_{j+1}) - W(t_j) $. $ \Delta W_j $ is the increment of the Wiener process or a white noise.\par
We next assume that the interacting-traders' excess demand for the index is approximated as $ Q^I_j = b\; n\; X_j $. Following [17], we assume that the interaction strengths change randomly in time $j$, and furthermore, define the randomness as $ \phi_j = \rho \Delta W_j $, where the parameter $ \rho $ is a constant, and $\Delta W_j$ is the increment of the Wiener process. 

\subsection{Fundamentalists}
\noindent 
Fundamentalists are assumed to have a reasonable knowledge of the fundamental value of the index $S^*_j$. If the price $S_j$ is below the fundamental value $ S^*_j $, a fundamentalist tends to buy the index (as he estimates the index to be undervalued), and if the price is above the fundamental value, a fundamentalist tends to sell the index. Hence we assume that fundamentalists' buying or selling order is given by $ Q^f_j = a\; m\; \left(\ln S^*_j - \ln S_j \right) $, where $m$ is the number of fundamentalists, and $a$ parametrizes the strength of the reaction on the discrepancy between the fundamental price and the market price.

\subsection{Market price}
\noindent 
Let us leave the traders' decision-making processes and turn to the determination of the market price. We assume the existence of a market clearing system. In the system, a \textit{market maker} mediates the trading and adjusts the market price to the market clearing values. The market transaction is executed when the buying orders are equal to the selling orders. The balance of demand and supply is written as $  Q^f_j + Q^I_j = a\; m\; \left[\ln S^*_j - \ln S_j \right] + b\; n\; X_j = 0 $. 
The market price is therefore calculated as $ \ln S_j = \ln S^*_j + \lambda X_j $ where  $ \lambda = \frac{b\; n}{a\; m} $. Using the price equation, we can categorize the market situations as follows: If $ X_j = 0 $, the market price $ S_j $ is equal to the fundamental price $ S^*_j $. If $ X_j > 0 $, the market price $ S_j $ exceeds the fundamental price $ S^*_j $ ({\it bull} market regime). If $ X_j < 0 $, the market price $ S_j $ is less than the fundamental price $ S^*_j $ ({\it bear} market regime). Using the price equation, the log-return is defined as 
\begin{equation}
 R_j = \ln S_{j} - \ln S_{j-1} = (\ln S^*_j - \ln S^*_{j-1}) + \lambda (X_j - X_{j-1}). 
%(15)
\end{equation}
For simplicity of analysis we here assume that the fundamental price $ S^*_j $ remains constant. 

\section{Simulations}
\noindent 
We are interested in the statistical properties of time series generated by our model and how they compare with those of real data shown in Section 2. In this section we perform numerical simulation of the model introduced in Section 3.\par
To this end, we consider two cases with different sets of parameters: 
\begin{enumerate} 
\item Case I: $ n = 100000 $; $ m= 10000 $; $ a = b = 1 $; $\lambda = 10 $; $\rho = 5 $; $ \Delta t_j = 0.1 $. 
\item Case II: $ n = 1000 $; $ m= 10000 $; $ a = b =1 $; $\lambda = 0.1$; $\rho = 2 $; $ \Delta t_j = 0.1 $. 
\end{enumerate}
Case I describes the situation of the stock market in the inflationary period. Namely, 
Case I assumes that (i) a large number of traders participate in trade, and (ii) the number of interacting traders is greater than the number of fundamentalists, and (iii) the interaction among the interacting traders is strong. In contrast, Case II describes the situation of the stock market in the deflationary period. Case II assumes that (i) a small number of traders participates in trade, and (ii) the number of interacting traders is less than the number of fundamentalists, and (iii) the interaction among the interacting traders is weak\footnote{For the mathematical explanation and results of the comprehensively numerical simulations of this model, see [20].}.\par
Figure 4(a) displays a typical time series of the absolute log-returns generated numerically from the model with Case I. Figure 4(b) shows the complementary cumulative distribution function in log-log scale, calculated from the model series of the absolute log-returns in Fig. 4(a). The figure shows that the complementary cumulative distribution function of the absolute log-returns is approximated by the power-law distribution $ P(V \geq x) \sim x^{-\alpha} $with the exponent $ \alpha $ equal to $1.7$.\par 
We next perform the numerical simulation of the model with Case II. Figure 5(a) shows a typical time series of the absolute log-returns generated from the model with Case II. Figure 5(b) indicates the semi-log plot of the complementary cumulative distribution function of the absolute log-returns calculated using the model series of the absolute log-returns in Fig. 5(a). The figure shows that the complementary cumulative distribution function of the absolute log-returns is accurately described by the exponential distribution $ P(V \geq x) \sim exp(x/\beta) $. The estimated parameter $\beta$ is equal to $0.7$. \par

\section{Concluding remarks}
\noindent 
In this paper we first showed the statistical properties of the Nikkei 225 index, that is, (i) in periods of inflation the complementary cumulative distribution function of the absolute log-returns is approximated by the power-law distribution, and (ii) in periods of deflation the complementary cumulative distribution function of the absolute log-returns is accurately described by the exponential distribution. \par
We next introduced a stochastically macroscopic model of stock markets to explain these empirical findings. In our model, the stock market is composed of two groups of traders: the fundamentalists and the interacting traders. The results of the numerical simulation suggest the following: in the period of inflation, the number of interacting traders strongly influenced by the investment attitudes of others is greater than the number of fundamentalists, who believe that the asset price will return to the fundamental price, so that power-law distribution of the absolute log-returns is generated by interacting traders' herd behavior. On the other hand, we can deduce that in the period of deflation a number of interacting traders leave the stock markets, and furthermore the interaction among the interacting traders is weak, while most of fundamentalists stay in the market; as such, the exponential law of absolute log-returns is observed. \par
As a consequence, the model captures the two opposing empirical data sets observed in real financial markets, and provides a reasonable explanation of the empirical findings we presented in section 2. Although the model is simple, it may thus be seen as a reasonable approximation of the absolute log-returns in real stock markets. 

\section{Acknowledgements}
\noindent
I wish to express my gratitude to Michiyo Kaizoji for her helpful discussion and her assistance in collecting the data, and Enrico Scalas for helpful suggestions and comments. 
This research was supported in part by the Japan Society for the Promotion of Science (JSPS) under the Grant-in-Aid, No.06632.

%------------------------------------------
%  Figure captions                     
%------------------------------------------
\begin{figure}[htbp] 
\begin{center}
  \includegraphics[height=19cm,width=14cm]{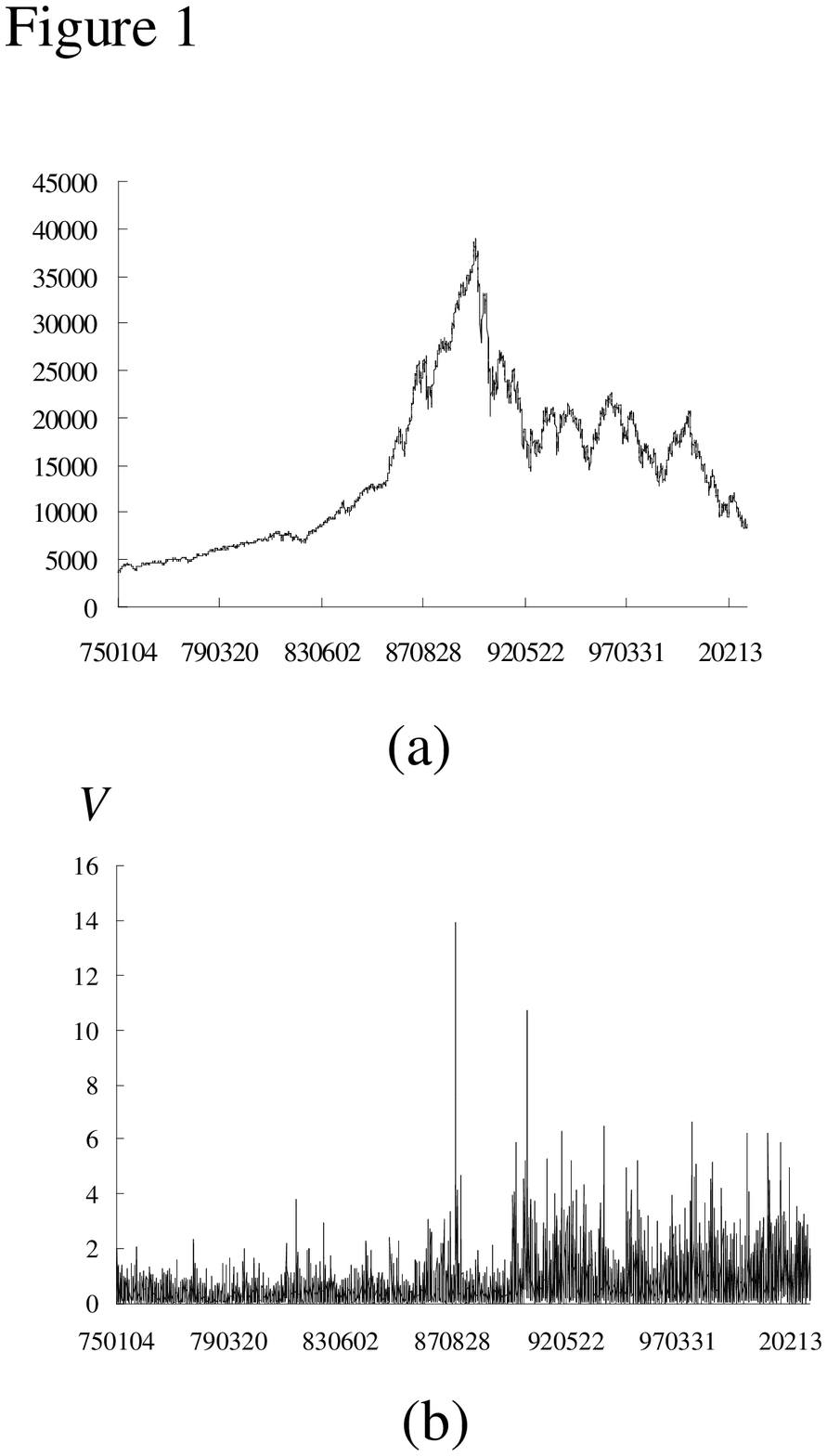}
\end{center}
\caption{(a) The time series of the Nikkei 225 index from January 1975 to December 2002, and (b) The time series of the volatility for the Nikkei 225 index from January 1975 to December 2002.}
\label{fig1.eps}
\end{figure}

\begin{figure}[htbp] 
\begin{center}
  \includegraphics[height=19cm,width=14cm]{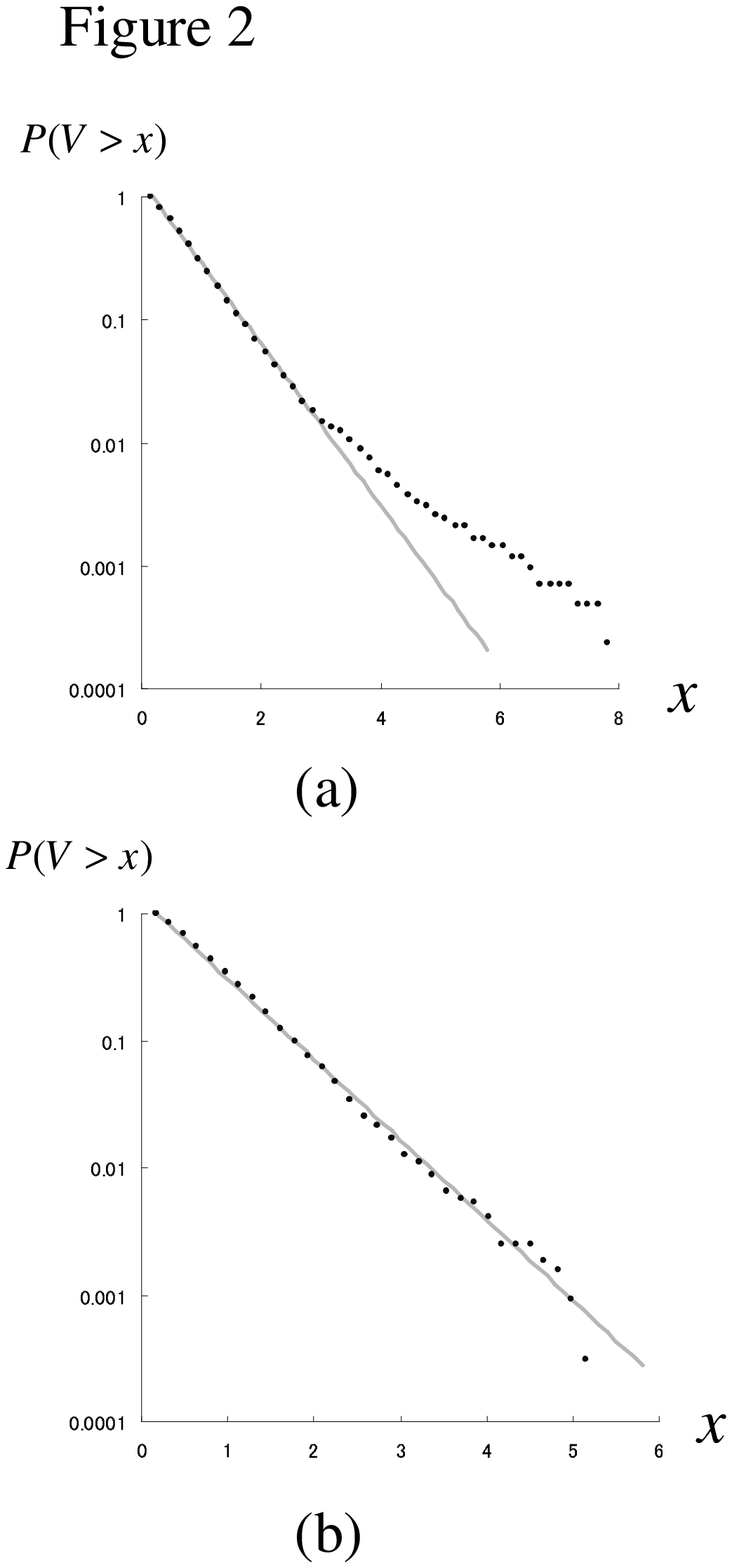}
\end{center}
\caption{(a) The complementary cumulative distribution function of the absolute log-returns for the Nikkei 225 index in the inflationary period (January 1975 - December 1989) shown in semi-log coordinates. Points: the actual data. Solid lines: fits to the exponential function. The temperature is equal to $ \beta = 0.66 $. (b) The complementary cumulative distribution function of the absolute log-returns for the Nikkei 225 index in the deflationary period (January 1990 - December 2002) shown in semi-log coordinates. Points: the actual data. Solid lines: fits to the exponential function. The temperature is equal to $ \beta = 0.68 $.}
\end{figure}

\begin{figure}[htbp] 
\begin{center}
  \includegraphics[height=19cm,width=14cm]{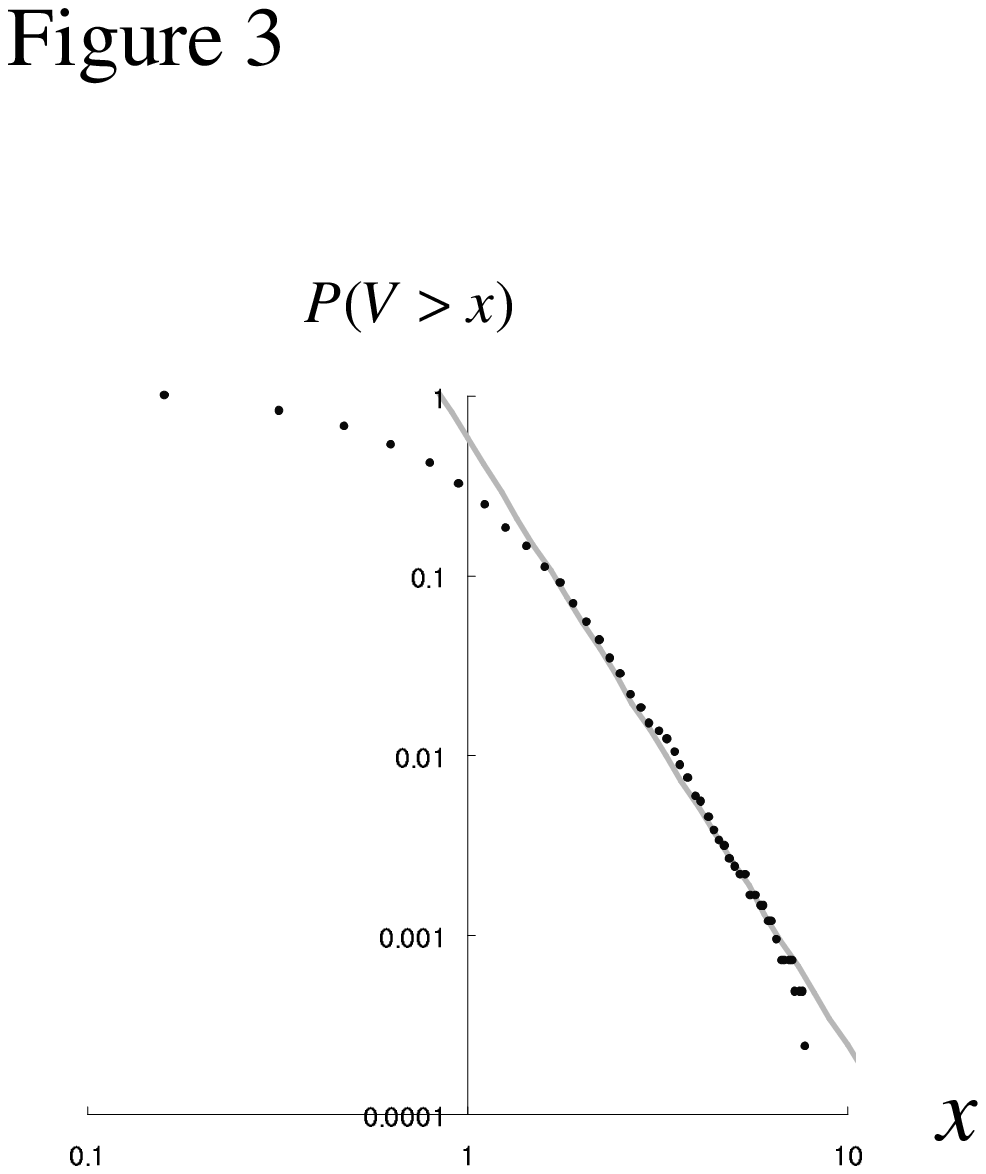}
\end{center}
\caption{The complementary cumulative distribution function of the absolute log-returns for the Nikkei 225 index in the inflationary period (January 1975 - December 1989) shown in log-log coordinates. Points: the actual data. Solid lines: fits to the power-law function. The power-law exponent is equal to $ \alpha = 3.38 $.}
\end{figure}

\begin{figure}[htbp] 
\begin{center}
  \includegraphics[height=19cm,width=14cm]{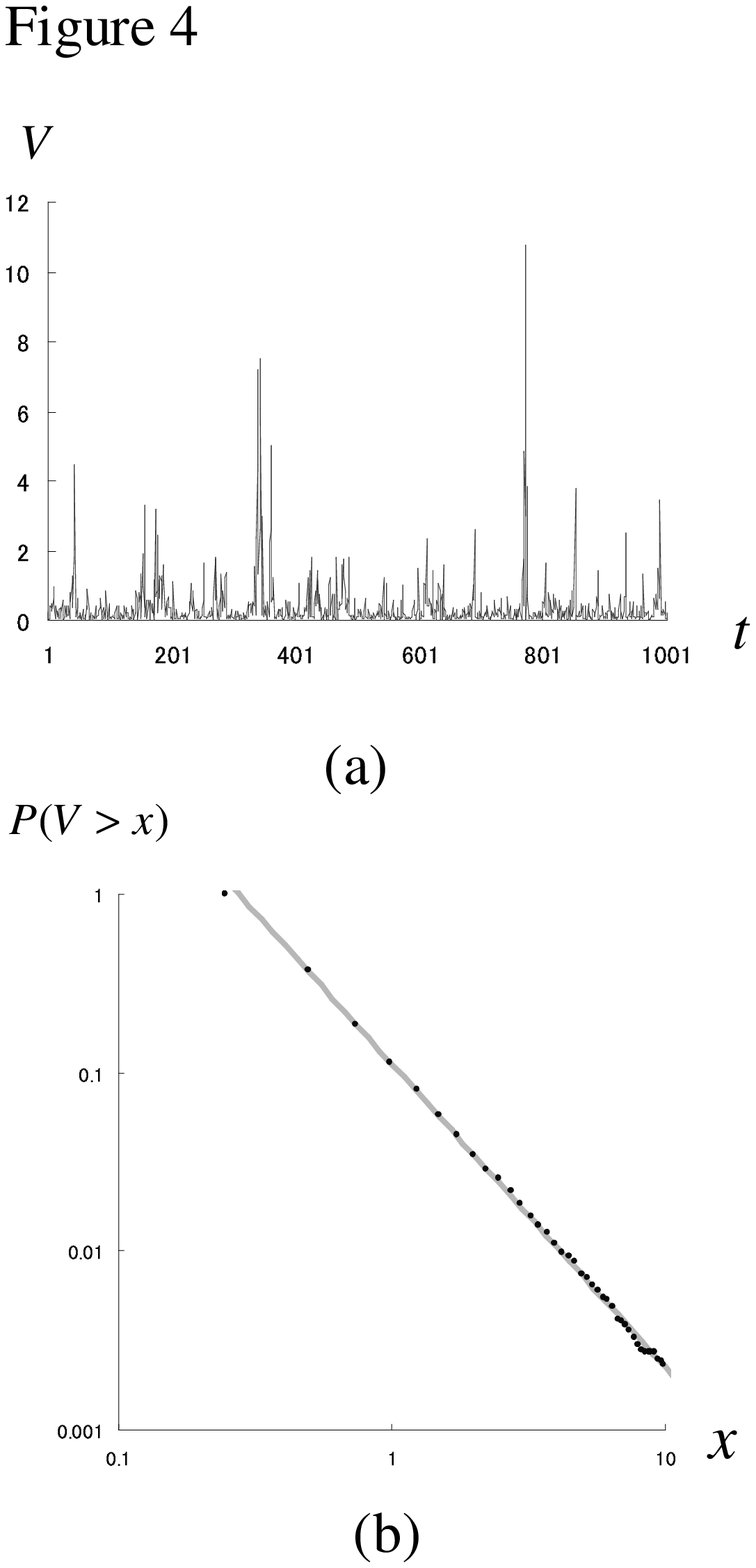}
\end{center}
\caption{(a) Model series of the absolute log-returns. The series are generated from the model with Case I, which includes the following set of parameters: $ n = 100000 $; $ m= 10000 $; $ a = b = 1 $; $\lambda = 10 $; $\rho = 5 $; and $\Delta t_j = 0.1$. (b) The log-log plot of the complementary cumulative distribution function of the absolute log-returns generated from the series in Fig. 4(a). Solid lines: fits to the power-law function with the exponent $ \alpha = 1.7 $.} 
\end{figure}

\begin{figure}[htbp] 
\begin{center}
  \includegraphics[height=19cm,width=14cm]{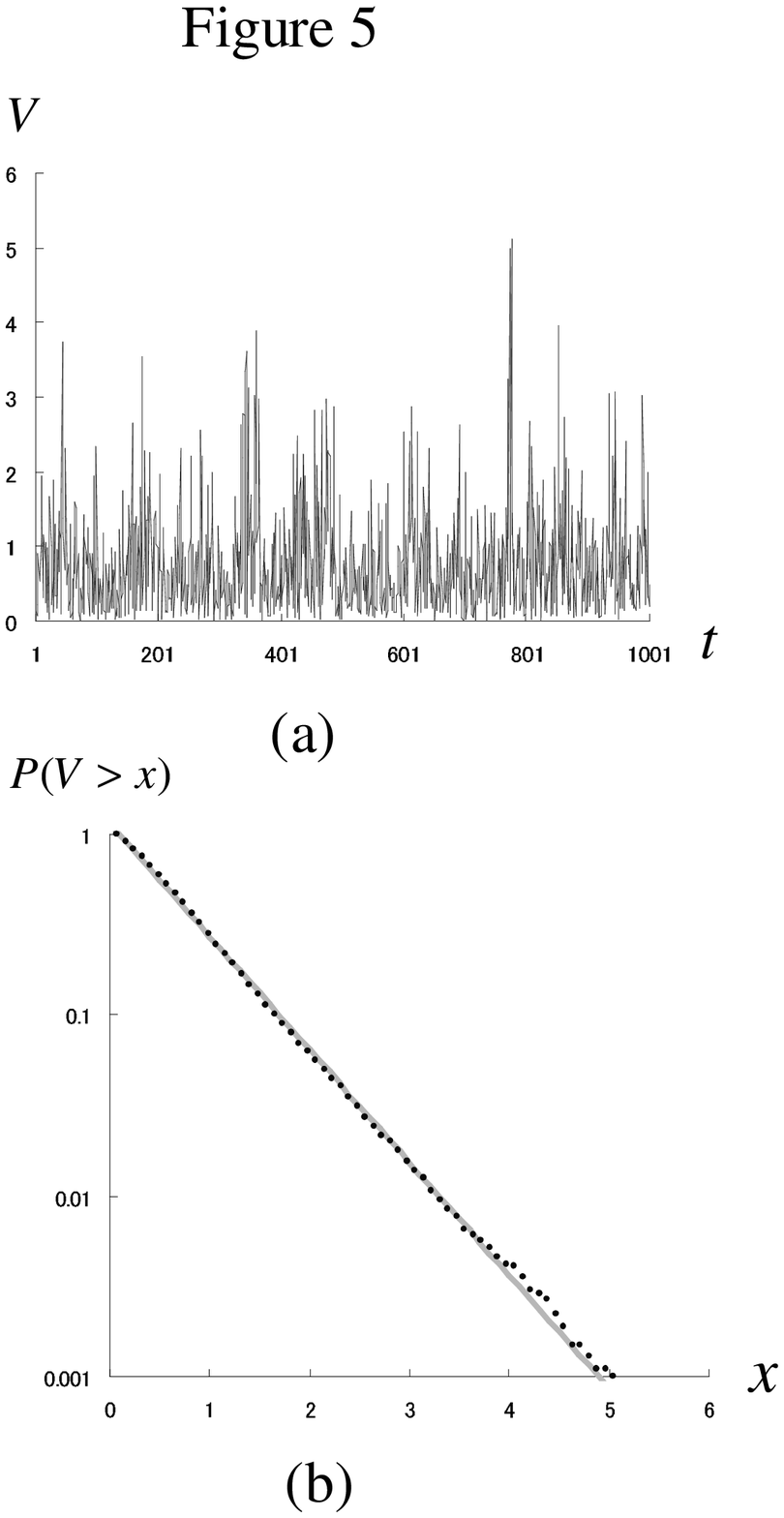}
\end{center}
\caption{(a) Model series of the absolute log-retunrs. The series are generated from the model with Case II, which includes the following set of parameters: $ n = 1000 $; $ m= 10000 $; $ a = b = 1 $; $\lambda = 0.1 $; $\rho = 2 $; and $\Delta t_j = 0.1$. (b) The semi-log plot of the complementary cumulative distribution function of the absolute log-returns generated from the series in Fig. 5(a). Solid lines: fits to an exponential function with the temperature $ \beta = 0.7 $. } 
\end{figure}
\end{document}